\documentstyle[12pt,epsfig]{article}                                                    

\newlength{\dinwidth}                                                    
\newlength{\dinmargin}                                                    
\def\lapproxeq{\lower .7ex\hbox{$\;\stackrel{\textstyle                                                    
<}{\sim}\;$}}                                                    
\def\gapproxeq{\lower .7ex\hbox{$\;\stackrel{\textstyle                                                    
>}{\sim}\;$}}                                                    
\def\be{\begin{equation}}                                                    
\def\ee{\end{equation}}                                                    
\def\bea{\begin{eqnarray}}                                                    
\def\eea{\end{eqnarray}}                                                    
\def\GeV{\rm GeV}
                                                    
\def\funp{{I\!\!P}}                    
                                                    
\begin{document}                                                    
\titlepage                                                    
\begin{flushright}                                                    
IPPP/08/78   \\
DCPT/08/156 \\                                                    
\today \\                                                    
\end{flushright}                                                    
                                                    
\vspace*{2cm}                                                    
                                                    
\begin{center}                                                    
{\Large \bf Checking formalism for central exclusive production in the first LHC runs\footnote{Based on a talk by A.D. Martin at the CERN - DESY Workshop "HERA and the LHC", 26 - 30 May 2008, CERN.}}                                                    
                                                    
\vspace*{1cm}                                                    
V.A. Khoze$^{a,b}$, A.D. Martin$^a$ and M.G. Ryskin$^{a,b}$ \\                                                    
                                                   
\vspace*{0.5cm}                                                    
$^a$ Institute for Particle Physics Phenomenology, University of Durham, Durham, DH1 3LE \\                                                   
$^b$ Petersburg Nuclear Physics Institute, Gatchina, St.~Petersburg, 188300, Russia            
\end{center}                                                    
                                                    
\vspace*{2cm}                                                    
\begin{abstract}

We discuss how the early LHC data runs can provide
crucial tests of  the 
formalism used to predict the cross sections of central exclusive production.
\end{abstract}

\section{Introduction}
The physics potential of forward proton
tagging at the LHC has attracted much attention in the
last years, for instance, \cite{INC}~-~\cite{pb}. 
The combined detection of both outgoing protons and the centrally
produced system gives access to a unique rich programme
of studies in QCD, electroweak and BSM physics. Importantly,
 these measurements will 
 provide valuable information on the Higgs sector of MSSM and other popular
BSM scenarios, see \cite{KKMRext}~-~\cite{fghpp}. 
%In particular, the CEP
%process allows for the
%unique possibility to study directly the coupling of
% Higgs-like bosons to bottom quarks
%\cite{INC,KMRmm}.

\begin{figure}[ht]
\begin{minipage}[t]{0.25\linewidth}
\centering
\includegraphics[height=4cm]{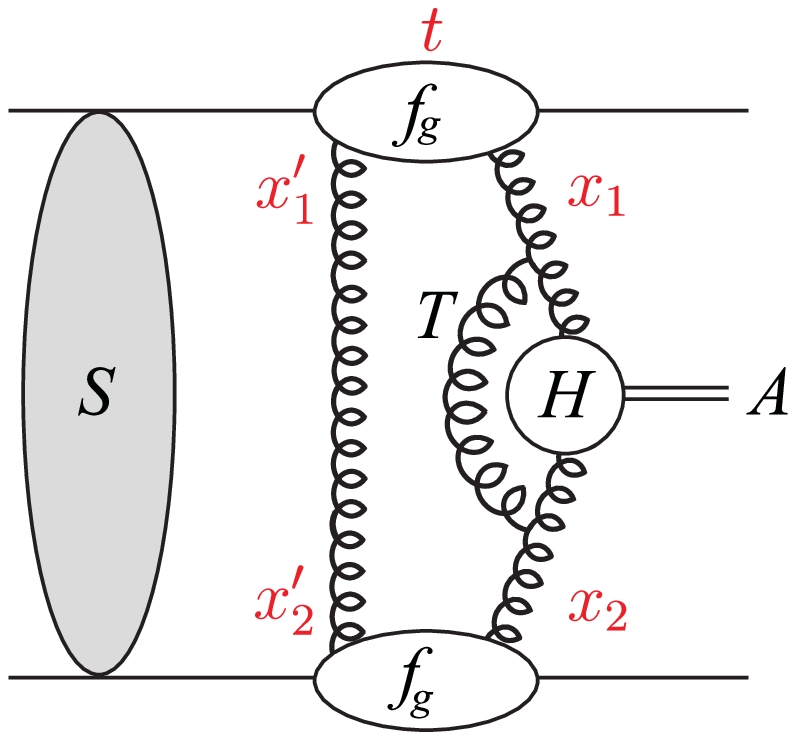}
\caption{A symbolic diagram for the CEP of a system  $A$.}
\label{fig:parts}
\end{minipage}
\hspace{0.5cm}
\begin{minipage}[t]{0.70\linewidth}
\centering
\includegraphics[height=4.5cm]{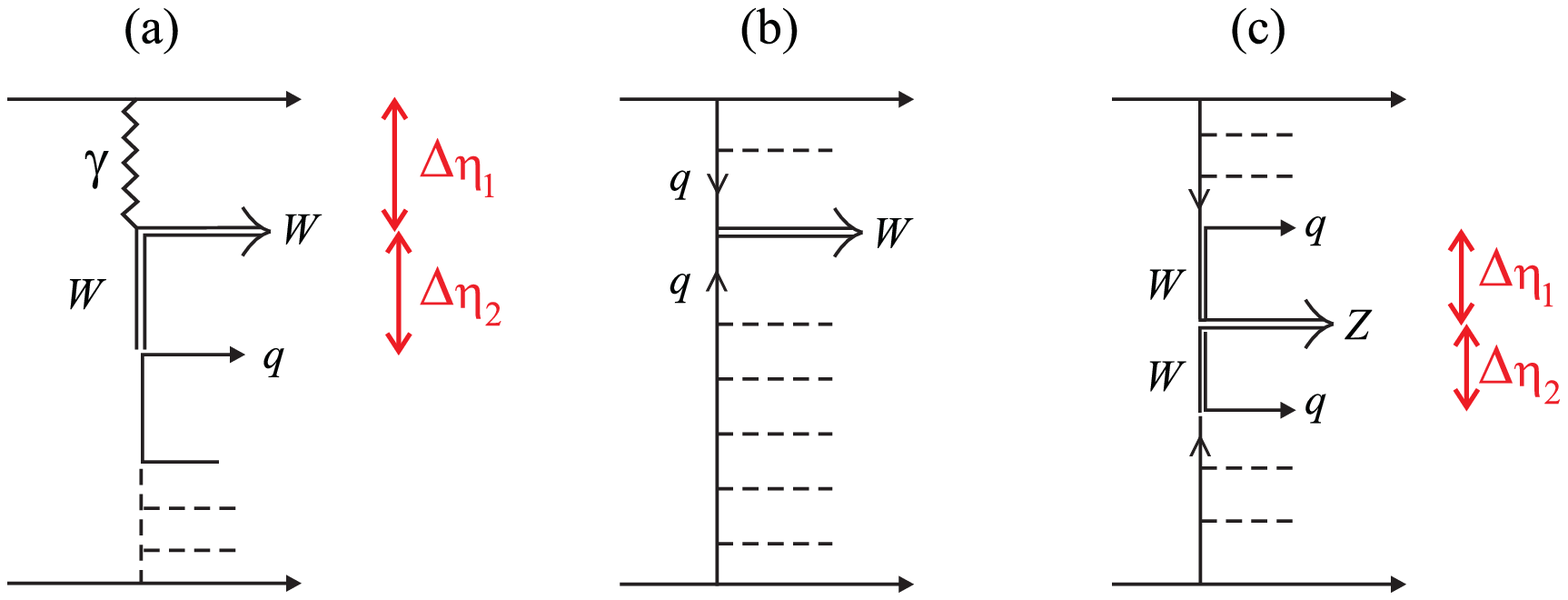}
\caption{(a) $W$ production with 2 gaps, (b) Inclusive $W$ production, (c) $Z$ production with 2 gaps.}
\label{fig:WZ}
\end{minipage}
\end{figure}

The theoretical formalism \cite{KMR}~-~\cite{KMRnewsoft}
% \cite{KMR,INC,KMRsoft,KMRnewsoft}
 for the description of a central exclusive production
(CEP) process contains quite distinct parts, shown symbolically
in Fig.~\ref{fig:parts}.
We first have to calculate the $gg \to A$ subprocess, $H$, convoluted with the gluon distributions $f_g$.
 Next, we must account for the QCD corrections 
 which reflect the absence of additional radiation in the hard subprocess -- that is,
  for the Sudakov factor $T$. Finally, we must enter soft physics to calculate
 the survival probability $S^2$ of the rapidity gaps (RG) .

%\begin{figure}
%\begin{center}
%\includegraphics[height=3cm]{parts.eps}
%\caption{A symbolic diagram for the central exclusive production of a system $A$.}
%\label{fig:parts}
%\end{center}
%\end{figure}

The uncertainties of the CEP predictions
are potentially not small. Therefore,
 it is important to perform checks using processes  that will be accessible in the first 
  LHC runs \cite{KMRearly}.
%with integrated luminosities in the range 100 ${\rm pb}^{-1}$ to 1 ${\rm fb}^{-1}$.
%In \cite{KMRearly} the reactions were
%identified,  where the different ingredients of the CEP formalism 
%could be tested experimentally.
We first consider measurements which do not rely on proton tagging and 
%Without tagging, diffractive measurements at the LHC
can be performed through the detection of RG. 

The main uncertainties of the CEP predictions  are associated with 
\begin{itemize}
\item [(i)] the probability $S^2$ that additional secondaries will not populate the gaps; 

\item [(ii)] the probability to find the appropriate gluons, that are given by generalized,
 unintegrated distributions $f_g(x,x',Q_t^2)$;
\item [(iii)] the higher order QCD corrections to the hard subprocess, in particular, the  Sudakov suppression; 
\item [(iv)] the so-called semi-enhanced absorptive corrections (see \cite{KKMR,bbkm}) 
and other effects, which may violate the soft-hard factorization.
\end{itemize}

%We discuss below, in turn, possible checks of the various ingredients of the calculation of the CEP cross sections. 
%\begin{figure}
%\begin{center}
%\includegraphics[height=3cm]{WZ.eps}
%\caption{(a) $W$ production with 2 gaps, (b) Inclusive $W$ production, (c) $Z$ production with 2 gaps.}
%\label{fig:WZ}
%\end{center}
%\end{figure}

\section{Gap survival factor $S^2$}

Usually, the gap survival  
is calculated within a multichannel eikonal approach \cite{GLMrev}. 
The probability $S^2$ of elastic $pp$ rescattering, 
shown symbolically  by $S$ in 
Fig.~\ref{fig:parts} 
can be evaluated in a model independent way once  the 
elastic cross section $d\sigma_{\rm el}/dt$ is measured at the LHC.
 However, there may be  excited states  between the blob $S$ and the 
  amplitude on the r.h.s of Fig.~\ref{fig:parts}.
  The presence of such states enlarges absorption. 
To check experimentally  the role of this effect,
    we need  a process with 
 a bare cross section that can be reliably calculated.
 Good candidates are the production of $W$ or $Z$ bosons with RGs \cite{KMRearly}.
%\subsection{$W$ production with rapidity gaps} 
%\begin{figure}
%\begin{center}
%\includegraphics[height=3cm]{WZ.eps}
%\caption{(a) $W$ production with 2 gaps, (b) Inclusive $W$ production, (c) $Z$ production with 2 gaps.}
%\label{fig:WZ}
%\end{center}
%\end{figure}
In the case of `$W$+gaps' production the main contribution comes from the diagram of
 Fig.~\ref{fig:WZ}(a) 
\cite{KMRphoton}. One gap, $\Delta \eta_1$, is associated with photon exchange, 
while the other, $\Delta \eta_2$, is associated with the $W$.
In the early LHC data runs the ratio ($W$+gaps/$W$ inclusive) will be measured first.
This measurement is a useful check of the models for soft rescattering \cite{KMRearly}.
%\subsection{$Z$ production with RG} 

A good way 
to study the low impact parameter ($b_t$) region 
 is to observe $Z$ boson production via $WW$ fusion, see Fig.~\ref{fig:WZ}(c). Here, both gaps
 originate from $W$-exchange, and the corresponding $b_t$ region 
 is similar to that for exclusive Higgs production. 
 The expected $Z$+gaps cross section is of the order of 0.2 pb,
 and $S^2$=0.3 for $\Delta \eta_{1,2} > 3$ and for quark jets with $E_T>50$ GeV \cite{pw}. 

%One problem is that even with the $E_T>50$ GeV cut, the QCD background arising
% from  QCD $b\bar{b}$ central exclusive production is comparable to the electroweak 
%$q\bar{q} \to Z+2$ jet %signal. Therefore, we should concentrate on the leptonic decay modes
% of the $Z$ boson, %which results in a smaller event rate
%\footnote{Note that in the recent study \cite{nikit}
% it was demonstrated that the so-called Track Counting Veto %(TCV) is robust for selection of the central
%RG events in vector-boson fusion 
%searches at CMS.}.

% \subsection With tagged protons.

%So far we have discussed measurements which do not depend on observing the forward protons. 
%However, 
%When the forward proton detectors become operational we can do more.
%In particular, we can 
%Both the longitudinal and transverse momentum of the protons can be measured, and we
%we can study the $k_t$ behaviour of the cross section sections  and 
%scan the proton opacity \cite{KMRphoton}.

\section{Generalized, unintegrated gluon distribution $f_g$}
The cross section for the CEP of a system $A$ 
essentially has the form \cite{KMR}
\begin{equation}
\sigma(pp \to p+A+p) ~\simeq ~\frac{S^2}{B^2} \left|\frac{\pi}{8} \int\frac{dQ^2_t}{Q^4_t}\: f_g(x_1, x_1', Q_t^2, \mu^2)f_g(x_2,x_2',Q_t^2,\mu^2)~ \right| ^2~\hat{\sigma}(gg \to A).
\label{eq:M}
\end{equation}
Here the factor $1/B^2$ arises from the integration over the proton transverse momentum. 
 Also, $f_g$ denotes the generalized, unintegrated gluon distribution.
% in the limit of $p_t \to 0$. 
In our case the distribution $f_g$
% has not yet been measured explicitly.
% However, in our case it 
can be obtained from the conventional gluon distribution, $g$,
  known from the global parton analyses.
The main uncertainty here comes from the lack of knowledge 
of the integrated gluon distribution $g(x,Q_t^2)$ at low $x$ and small scales.
 For example, taking $Q_t^2=4~\GeV^2$ we find \cite{KMRearly}
% that a variety of recent MRST \cite{mstw} and CTEQ \cite{cteq} global analyses give a spread of
%\be
$xg~=~(3-3.8)~~{\rm for}~~x=10^{-2}~~~~{\rm and}~~~~xg~=~(3.4-4.5)~~{\rm for}~~x=10^{-3}$.
%\ee
These are big uncertainties bearing in mind that the CEP cross section 
 depends on $(xg)^4$.
%\subsection{Exclusive $\Upsilon$ production as a probe of $f_g$}
\begin{figure} [t]
\begin{center}
\includegraphics[height=5cm]{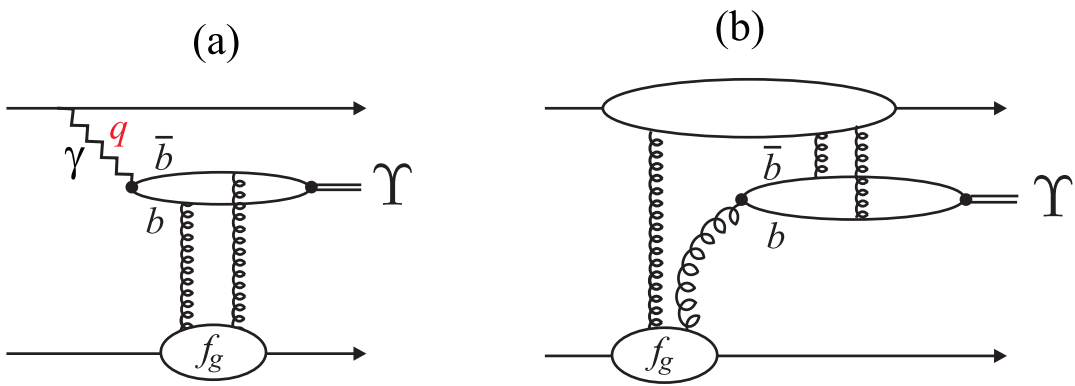}
\caption{Exclusive $\Upsilon$ production via (a) photon exchange, and (b) via odderon exchange.}
\label{fig:upsilon}
\end{center}
\end{figure}
To reduce the uncertainty associated with $f_g$ we can measure exclusive $\Upsilon$ production.
%\footnote{
%A feasibility study of the $\gamma p \to \Upsilon p$ measurement performed by CMS \cite{YCMS}
%looks quite encouraging.}.
 The process is shown in Fig.~\ref{fig:upsilon}(a). 
  The cross section for $\gamma p \to \Upsilon p$ is given in terms
   of the same  unintegrated gluon distribution $f_g$ that occurs in Fig.~\ref{fig:parts}.
% If the $\Upsilon$ is detected at the LHC in the rapidity interval $(|y|< 2.5)$ then it will sample the $x_{1,2}$ %intervals ($10^{-4}-10^{-2}$). 
There may be competition between production via photon exchange, 
Fig.~\ref{fig:upsilon}(a),
 and via odderon exchange,
%, where the upper three $t$-channel gluons form the odderon state,
 see Fig.~\ref{fig:upsilon}(b).
% To date, odderon exchange has not been observed.
A lowest-order  calculation (e.g.\cite{bmsc} )
 indicates that the odderon process (b) may be comparable to the photon-initiated process (a). 
 If the upper proton is tagged, it will be straightforward to separate the two mechanisms.
%since odderon production has no $1/q^2$.
%singularity characteristic of the photon. 
%The expression for $\sigma(\gamma p \to \Upsilon p) \propto f_g^2$ is given in \cite{mrt}. 
%In order to use this process to constrain the gluon distribution
%it would be preferable to tag the lower proton.

\section{Three-jet events as a probe of the Sudakov factor}

The search for the exclusive dijets at the 
Tevatron, $p\bar{p} \to p+jj+\bar{p}$, is performed \cite {CDFjj} by plotting 
the cross section in terms of the variable
%\be
$R_{jj}=M_{jj}/M_A$, where $M_A$ is the mass of the whole central system.
%\label{eq:jj}
%\ee
%Ideally, we might expect exclusive dijet production to show up as a narrow peak centred at $R_{jj}=1$.
However, 
the $R_{jj}$  distribution is smeared out by
QCD radiation, hadronization, the jet algorithm and 
 other experimental effects \cite {CDFjj,KMRrj}.
%
%
%\begin{figure}
%\begin{center}
%\includegraphics[height=7cm]{f2.eps}
%\caption{The rapidities of the three jets in the central system. Note that the rapidity $y_A$ of the whole %central system does not necessarily occur at $y=0$. The rapidity
%interval containing the three jets is denoted by $\delta\eta$, outside of which there is no hadronic
%activity.\label{fig:2}}
%\end{center}
%\end{figure}
To weaken the smearing it was proposed in Ref.~\cite{KMRrj} 
to study the  dijets  in terms of a  variable
%\be
$R_j~=~2E_T ~({\rm cosh}~\eta^*)/M_A~$,
%\label{eq:j}
%\ee
where only the transverse energy and the rapidity $\eta$ of the jet with
the {\it largest} $E_T$ enter.
 Here $\eta^* = \eta -y_A$, where $y_A$ is the rapidity of the  central system.
Clearly,
the largest $E_T$ jet is less affected by the smearing.  
As shown in \cite{KMRearly}, it is sufficient to consider
the emission of a third  jet,
% as shown in Fig.~\ref{fig:2},
when we take all three jets to lie in a specified rapidity interval $\delta\eta$. 
The cross section $d\sigma/dR_j$, as a function of $R_j$, for the production of a
pair of  high $E_T$ dijets accompanied by a third  jet is
 discussed in \cite{KMRrj,KMRearly}. 
It is shown that 
 the measurements of the exclusive two- and three-jet cross sections {\it as a function of $E_T$}
  of the highest jet allow a detailed check of the Sudakov physics; with much more information coming from
 the $\delta\eta$ dependence study. 
A clear way to observe the Sudakov suppression is just to measure the $E_T$
 dependence of exclusive dijet production. On dimensional grounds
  we would expect $d\sigma/dE_T^2 \propto 1/E_T^4$. 
  This behaviour is modified by the gluon anomalous dimension and by a stronger 
  Sudakov suppression with increasing $E_T$. Already the existing CDF 
   dijet data \cite{CDFjj} exclude predictions which omit the Sudakov effect.
\section{Soft-hard factorization: enhanced absorptive effects}
\begin{figure} 
\begin{center}
\includegraphics[height=6cm]{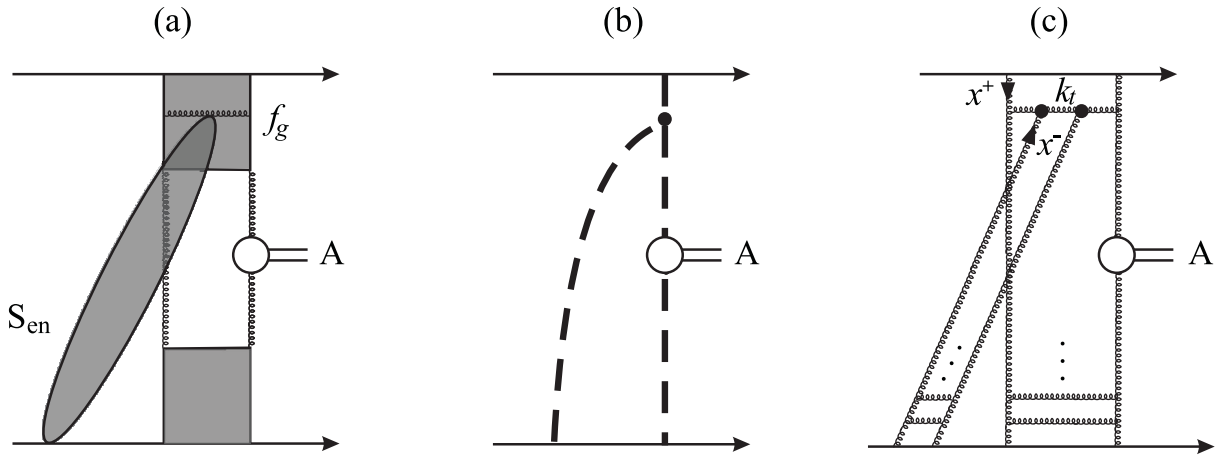}
\caption{(a) A typical enhanced diagram, where the shaded boxes denote $f_g$,
 and the soft rescattering is on an intermediate parton,
 giving rise to a survival factor $S_{\rm en}$; (b) and (c) are the Reggeon and QCD 
 representations, respectively.}
\label{fig:enh}
\end{center}
\end{figure}
The soft-hard factorization implied by Fig.~\ref{fig:parts}
could be violated by the so-called enhanced Reggeon diagrams,
see  Fig.~\ref{fig:enh}(a).
%caused by the rescattering of an intermediate parton
% generated in the evolution of $f_g$. 
%Such a diagram is shown in Fig.~\ref{fig:enh}(a).
% Fig.~\ref{enh}(b) is the Pomeron representation of the diagram, where the rapidity of the intermediate parton fixes the %position of the triple-Pomeron vertex.
The contribution of the first Pomeron loop, Fig.~\ref{fig:enh}(b) 
was calculated in pQCD in Ref.~\cite{bbkm}.
 A typical  diagram is shown in Fig.~\ref{fig:enh}(c). 
 For LHC energies it was found that  such effect 
 may be numerically large.
The reason is that the gluon density grows at low $x$ and,
 for low $k_t$ partons, approaches the saturation limit.
  However, as discussed in \cite{KMRearly},     
the enhanced diagram should affect mainly the very 
beginning of the QCD evolution -- the region that cannot
 be described perturbatively and which, in  \cite{KMRsoft,KMRnewsoft},
is already included phenomenologically.

\begin{figure}[ht]
\begin{minipage}[b]{0.45\linewidth}
\centering
\includegraphics[height=8cm]{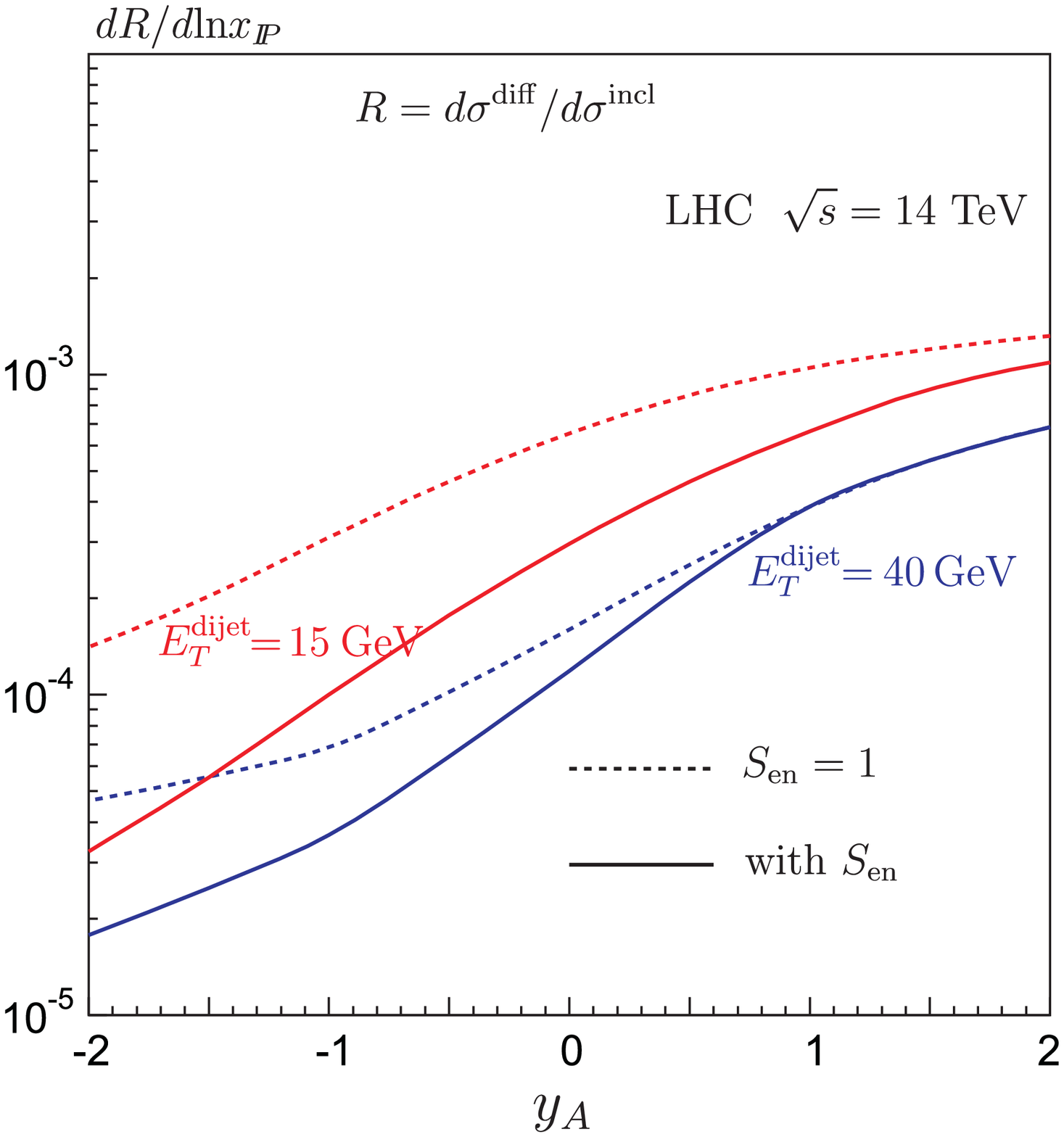}
\caption[The predictions]{The predictions of the ratio $R$ of (\ref{eq:Ren})
 for the production of a pair of high $E_T$ jets.} 
\label{fig:upsd}
\end{minipage}
\hspace{0.5cm}
\begin{minipage}[b]{0.5\linewidth}
\centering
\includegraphics[height=8.5cm]{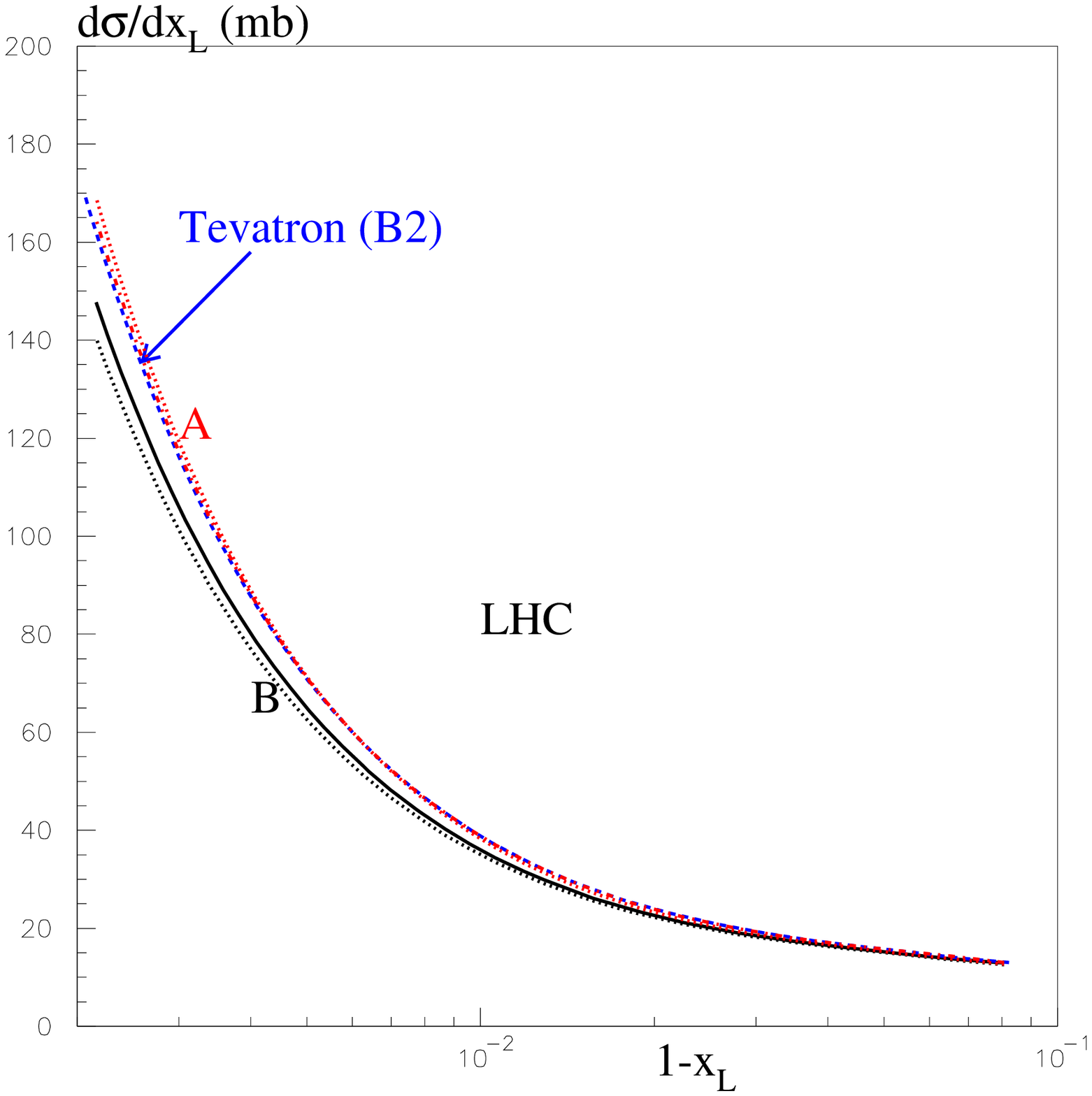}
\caption[The cross section]{The cross section $d\sigma_{\rm SD}/dx_L$ for single
 dissociation integrated over $t$ at the LHC
  energy.}
\label{fig:A2}
\end{minipage}
\end{figure}

Experimentally,  
we can study the role of semi-enhanced absorption  by measuring the
 ratio $R$ of diffractive event rate for $W$ (or $\Upsilon$ or dijet) 
 as compared to the inclusive process \cite{KMRearly}.
%\begin{figure} [t!]
%\begin{center}
%\centerline{\epsfxsize=0.7\textwidth\epsfbox{3en.eps}}
%\caption{\it Schematic diagrams for (a) the inclusive production of a system $A$, (b) and (c) for the diffractive production of $A$ without and with `enhanced' soft rescattering on intermediate partons. The system $A$ is taken to be either a $W$ boson or an $\Upsilon$ or a pair of high $E_T$ jets. .}
%\label{3en}
%\end{center}
%\end{figure}
%These processes are shown schematically in Fig.~\ref{3en}. In other words, $R$ is the ratio of the process 
%in diagram %(c) to that in diagram (a).
 That is
\be 
R~~=~~\frac{{\rm no.~of}~ (A+{\rm gap)~ events}}{{\rm no.~of~ (inclusive}~A)~{\rm events}}~~=~~
\frac{a^{\rm diff}(x_\funp ,\beta,\mu^2)}{a^{\rm incl}(x=\beta x_\funp,\mu^2)}~\langle S^2S^2_{\rm en}\rangle_{{\rm over}~b_t},
\label{eq:Ren}
\ee
where $a^{\rm incl}$ and $a^{\rm diff}$ are the parton
 densities determined from the global analyses 
 of inclusive and diffractive DIS data, respectively.
% The central system $A$ is either $W$ or a pair of high $E_T$ 
%jets or $\Upsilon$ or a Drell-Yan $\mu^+\mu^-$ pair.
% For $W$ or $\mu^+\mu^-$  production the parton densities $a$ are quark distributions, 
% whereas for dijet or $\Upsilon$ they are mainly gluon densities. 
% The measurements of the ratio $R$ will probe the gap survival factor averaged over the impact parameter $b_t$.
% we can measure the rapidity $y_A$ of the central system  and also the momentum fraction carried in the gap direction by %the soft hadrons, $\xi^- = \sum \xi_i^-$.
%Thus we know the value of $x_\funp=\xi^- +\xi^-_A$. That is, 
We can measure a double distribution $d^2 \sigma^{\rm diff}/dx_\funp dy_A$, and form the ratio
 $R$ using the inclusive cross section, $d\sigma^{\rm incl}/dy_A$.
 If we neglect the enhanced absorption, it is
quite straightforward to calculate the ratio $R$ of (\ref{eq:Ren}). 
The results for a dijet case 
are shown by the dashed curves in Fig.~\ref{fig:upsd}
  as a function of the rapidity $y_A$ of the 
dijet system.
The enhanced rescattering reduce the ratios
  and lead to steeper $y_A$ distributions, as illustrated by the continuous curves. 
%The plots correspond to a fixed value of $\xi^-=10~\GeV/7$ TeV. That is, we assume that the soft hadrons observed in the %central calorimeter carry a longitudinal momentum
% $p_z$=10 GeV in the direction of the RG associated with Pomeron exchange.
%
%
Perhaps the most informative probe of $S^2_{\rm en}$ 
is to observe the ratio $R$ for dijet production in the region $E_T \sim 15-30$ GeV.
 For example, for $E_T \sim$ 15 GeV 
 we expect $S^2_{\rm en} \sim$ 0.25, 0.4 and 0.8 at $y_A=-2, ~0$ and $2$ respectively. 
\section{Conclusion}
 
The addition of forward proton detectors to LHC experiments will
add unique capabilities to the existing LHC experimental programme.
For certain BSM scenarios, the tagged-proton mode may even be the
%Higgs 
discovery channel.
There is also a rich
QCD, electroweak, and 
more exotic physics, menu. 

The uncertainties in the prediction of the CEP processes 
are potentially not small. Therefore, it is crucial
 to perform checks of the theoretical formalism
using reactions that will be experimentally 
accessible in the first  LHC runs  \cite{KMRearly}.

Most of the  measurements 
discussed above can be performed, without detecting the  protons,
 by taking advantage of the relatively low luminosity in the early LHC  runs.
% This allows the use of a veto trigger to select events with  a large RG. 
% In this way we are able to probe the various  components of the theoretical
%  formalism.
When the forward proton detectors are operating much more can be done.
  First, it is possible to measure directly the cross section $d^2\sigma_{\rm SD}/dtdM^2_X$
 for single diffractive dissociation and also the cross section $d^2\sigma_{\rm DPE}/dy_1 dy_2$
 for soft central diffractive production. These measurements will strongly constrain the models 
 used to describe diffractive processes and the effects of soft rescattering. The 
recent predictions can be found in \cite{KMRnewsoft}. For illustration
we show in Fig. \ref{fig:A2} the expectation 
for $d\sigma_{\rm SD}/dx_L$, see for details \cite{KMRnewsoft}.
%\begin{figure}
%\begin{center}
%\includegraphics[height=5cm]{sdlhc.eps}
%\caption{The cross section $d\sigma_{\rm SD}/dx_L$ for single
% dissociation integrated over $t$ at the LHC 
% energy.}
%\label{fig:A2}
%\end{center}
%\end{figure}
Next, a study of the transverse momentum distributions of both of the tagged protons,
 and the correlations between their momenta, is able to scan the proton optical density
\cite{KMRphoton,KMRtag}.


\begin{thebibliography}{0}

\bibitem{INC} V.A.~Khoze, A.D.~Martin and M.G.~Ryskin, 
                   {\em Eur. Phys. J.} {\bf C23}, 311 (2002). 

\bibitem{DKMOR} A.~De~Roeck {\it et al.},
                {\em Eur. Phys. J.} {\bf C25}, 391(2002). 

\bibitem{Jeff} J.~Forshaw and A.~Pilkington,
  %``Central production of new physics,''
%\href{http://www.slac.stanford.edu/spires/find/hep/www?irn=7737734}{SPIRES entry}
{\it  In *Hamburg 2007, Blois07, Forward physics and QCD* 130-136}.

\bibitem{FP420}
  M.~G.~Albrow {\it et al.} [FP420 R and D Collaboration],
  %``The FP420 R&D Project: Higgs and New Physics with forward protons at the
  %LHC,''
  arXiv:0806.0302 [hep-ex].
  %%CITATION = ARXIV:0806.0302;%%

\bibitem{pb} P.~J.~Bussey,
  %``Detection of new states using forward proton tagging at the LHC,''
  arXiv:0809.1335 [hep-ex].
  %%CITATION = ARXIV:0809.1335;%%

%\bibitem{KMRlt} V.A.~Khoze, A.D.~Martin and M.G.~Ryskin,
%                arXiv:0705.2314 [hep-ph].

\bibitem{KKMRext} A.B.~Kaidalov {\it et al.},
                 {\em Eur. Phys. J.} {\bf C33}, 261 (2004).

\bibitem{HKRSTW} S.~Heinemeyer {\it et al.},{\em Eur.\ Phys.\ J.} {\bf C53}, 231 (2008).
%                    arXiv:0801.1974 [hep-ph]. 
\bibitem{CLP} B.~Cox, F.~Loebinger and A.~Pilkington, JHEP {\bf 0710}, 090 (2007).

\bibitem{fghpp}  J.~R.~Forshaw {\it et al.},
  %``Reinstating the 'no-lose' theorem for NMSSM Higgs discovery at the LHC,''
  JHEP {\bf 0804}, 090 (2008).
% [arXiv:0712.3510 [hep-ph]].
  %%CITATION = JHEPA,0804,090;%%
 
%P.~J.~Bussey, T.~D.~Coughlin, J.~R.~Forshaw and A.~D.~Pilkington,
%JHEP {\bf 0611}, 027 (2006);\\

%\bibitem{KMRmm} V.A.~Khoze, A.D.~Martin and M.~Ryskin, 
%                {\em Eur. Phys. J.} {\bf C19}, 477 (2001)
%[Erratum-ibid.\ {\bf C20}, 559 (2001)].

\bibitem{KMR} V.A.~Khoze, A.D.~Martin and M.G.~Ryskin, 
{\em Eur. Phys. J.} {\bf C14}, 525 (2000). 

\bibitem{KMRsoft} V.A. Khoze, A.D. Martin and M.G. Ryskin, {\em Eur. Phys. J.}  {\bf C18}, 167 (2000).

\bibitem{KMRnewsoft} M.G. Ryskin, A.D. Martin and V.A.~Khoze, {\em Eur. Phys. J.} {\bf C54}, 199 (2008);
E.G.S. Luna {\it et al.},arXiv:0807.4115 [hep-ph].
%``Diffractive dissociation re-visited for predictions at the LHC,''
 %%CITATION = ARXIV:0807.4115;%%


\bibitem{KMRearly} V.A.~Khoze, A.D.~Martin and M.G.~Ryskin,
 %``Early LHC measurements to check predictions for central exclusive
  %production,''
  Eur.\ Phys.\ J.\  C {\bf 55}, 363 (2008).
 %[arXiv:0802.0177 [hep-ph]].
  %%CITATION = EPHJA,C55,363;%%

                   
\bibitem{KKMR} A.B.~Kaidalov et al.,
  {\em Eur. Phys. J.} {\bf C21}, 521 (2001).
  
\bibitem{bbkm}J.~Bartels et al.,
                 Phys.\ Rev.\ {\bf D73}, 093004 (2006).
\bibitem{GLMrev} For a review see:
               E.~Gotsman et al.,
               arXiv:hep-ph/0511060.
% and references therein.
\bibitem{KMRphoton} V.A. Khoze, A.D. Martin and M.G. Ryskin, {\em Eur. Phys. J.} {\bf C24}, 459 (2002).  

\bibitem{pw} V.A. Khoze et al., {\em Eur. Phys. J.}  {\bf C26}, 429 (2003).

%\bibitem{nikit} M. Vazgues-Acosta et al.,
%to be published in the proceedings of Les Houches 2007.

%\bibitem{shuv} A.G. Shuvaev et al., Phys. Rev. {\bf D60} 014015 (1999). 

%\bibitem{mstw} A.D. Martin et al., Phys. Lett. {\bf B652}, 292 (2007) and references therein.

%\bibitem{cteq} W.-K. Tung et al., JHEP {\bf 0702}:053 (2007) and references therein.  

%\bibitem{cteq} W.-K. Tung et al., JHEP {\bf 0702}:053 (2007) and references therein.  

%\bibitem{mrt} A.D. Martin, M.G. Ryskin and T. Teubner, Phys.Lett. {\bf B454}, 339 (1999).

\bibitem{bmsc} A. Bzdak et al., Phys. Rev. {\bf D75}, 094023 (2007).

%\bibitem{YCMS}  J.~Hollar, S.~Ovyn and X.~Rouby, CMS note AN-2007/032.

\bibitem{CDFjj} 
 T.~Aaltonen {\it et al.}  [CDF Run II Collaboration],
  %``Observation of Exclusive Dijet Production at the Fermilab Tevatron p-pbar
  %Collider,''
  Phys.\ Rev.\  D {\bf 77}, 052004 (2008).
  % [arXiv:0712.0604 [hep-ex]].
  %%CITATION = PHRVA,D77,052004;%%
%K.Goulianos and J. Pinfold, these Proceedings.

\bibitem{KMRrj} V.A. Khoze, A.D. Martin and M.G. Ryskin, {\em Eur. Phys. J.} {\bf C48}, 467 (2006).  

%\bibitem{WZ} G. Watt, arXiv:0712.2670.

%\bibitem {KMRln} V.A. Khoze, A.D. Martin and M.G. Ryskin, JHEP {\bf 0605}:036 (2006).

\bibitem{KMRtag} V.A. Khoze, A.D. Martin and M.G. Ryskin, {\em Eur. Phys. J.}{\bf C24}, 581 (2002).



\end{thebibliography}
\end{document}